\documentclass[a4paper,11pt]{article}
\pdfoutput=1 

\usepackage{jinstpub} 

\title{\boldmath The PETALO project}

\author{Nerea Salor Iguiñiz}

\affiliation{Donostia International Physics Center (DIPC),\\Paseo Manuel Lardizabal 4, Donostia-San Sebastián, E-20018, Spain}

\emailAdd{nereasalor@gmail.com}

\abstract{PETALO (Positron Emission TOF Apparatus with Liquid xenOn) is a project that uses liquid xenon as a scintillation medium, silicon photomultipliers as a readout and fast electronics to provide a significant improvement in PET-TOF technology. Liquid xenon allows one to build a continuous detector with a high stopping power for 511 keV gammas. In addition, SiPMs enable a fast and accurate measurement of the time and energy with a small dark count rate at the low temperatures required by liquid xenon. PETit, the first PETALO prototype built at IFIC (Valencia), consists of an aluminum box with one volume of liquid xenon and two planes of VUV SiPMs, which register the scintillation light emitted in xenon by the gammas coming from a $^{22}$Na radioactive source placed in the middle. The liquid xenon volume is divided in small, highly reflective cells to enhance light collection.
The first results of energy resolution are presented. }


\collaboration[c]{on behalf of PETALO collaboration}

\proceeding{Light Detection in Noble Elements (LIDINE 2023)\\
  20-22 September 2023 \\
  School of Mines and Energy Engineering, Madrid (Spain)}

\begin{document}
\maketitle
\flushbottom

\section{Introduction}
\label{sec:intro}

Positron emission tomography (PET) is a technique used in hospitals to obtain images of the metabolic activity of the body. A molecule of glucose is modified, substituting an atom with a radioactive one, in this case a positron emission atom, like $^{18}$F. This glucose goes to the regions of the body with higher metabolic activity, such as brain and tumors, and it accumulates there. The positron emitted by the fludeoxyglucose  annihilates with one electron of the body of the patient, thus producing two photons of 511 keV in opposite directions. These photons can be detected in time coincidence if the patient is surrounded with a ring of photosensors. Detecting these photons allows one to draw a line connecting them, the line of response, and crossing lots of these lines allows one to reconstruct the image of the area where the photons are originated and so, where the metabolic activity of the patient is higher. 

In PET there are two important features, the energy resolution and the time resolution. Having a very good energy resolution allows one to reject scatter events which make the image blurred. In these events, at least one of the photons has interacted in the body of the patient, thus changing its direction and losing part of its energy. On the other hand, the time resolution allows one to obtain a time difference between the arrival of the two gammas in the detector. Thus, it is possible to identify a segment in the line of response with more probability of having emitted the photons, instead of taking into consideration the whole line in the reconstruction algorithm. As a consequence, the image quality can improve~\cite{a}  or an image with the same quality can be obtained using less radioisotope doses or reducing the time of the scans, which can be especially useful in pediatric oncology.

The PETALO project~\cite{b,c,d} consists of a PET based on liquid xenon instead of crystals as a scintillator material. Liquid xenon is a very good candidate to be used in PET due to its scintillation characteristics. On the one hand, it presents a high scintillation yield, as it produces around 68 photons per keV of energy deposited~\cite{e}, which leads to a better energy resolution. On the other hand, these photons emitted by the liquid xenon present a very fast decay time, of only 2.2 ns in its fastest component~\cite{f}. Consequently, there are lower fluctuations in the time determination, thus obtaining, in principle, a better time resolution. These scintillation photons are emitted in the ultraviolet range, at 178nm, and are detected using specific SiPMs sensitive to the VUV light. Furthermore, liquid xenon is a continuous material, allowing to have a uniform response in all the system, as it can continuously be purified passing through filters.

\section{First prototype: PETit}

PETit is the first prototype of PETALO and is operating at the Instituto de Física Corpuscular, in Valencia (Spain). It is an aluminum cube filled with liquid xenon and is hanging inside a vessel. To liquefy the xenon, a Sumitomo CH-110 cold head coupled to the cube via copper thermal links is used to reach the temperature of -110ºC. The vessel is kept in vacuum to maintain the xenon cryogenic temperature, as it reduces heat dissipation. The xenon inside the cube is continuously being recirculated in gas phase with a double diaphragm compressor and purified passing through a PS4 MT15 R2 hot getter filter from Sigma Technologies.This filter removes the impurities in the gas, such as oxygen, nitrogen and water, which can quench xenon scintillation light~\cite{g}. 

The prototype is instrumented with two planes of VUV SiPMs, one in front of the other. They have an active area of 6 x 6 mm$^2$ and they are arranged in arrays of 8 x 8 sensors each. The SiPMs have high gain and a fast response, which is necessary to obtain a good time resolution and they also present a low dark noise when they are working at cryogenic temperatures. A $^{22}$Na calibration source, which emits positrons, is placed in the middle of the cube, in a port, so that the source is not in contact with the liquid xenon. 

\begin{figure}[htbp]
\centering 
\includegraphics[width=1\textwidth,]{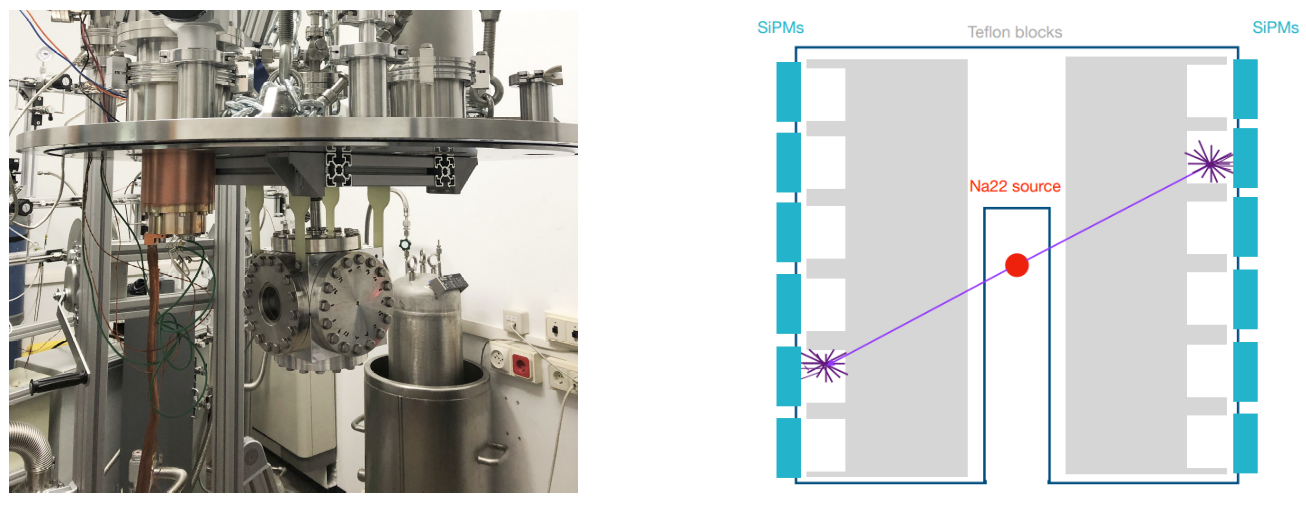}

\caption{\label{fig:sketch} Left: the experimental setup, with the cube containing the liquid xenon. Right: the inner sketch of PETit (not to scale).}
\end{figure}

The aim of PETit is to obtain the best possible results in energy and time resolution, for which it is necessary to enhance the light collection and minimize distortions introduced by the depth of interaction. To this aim, a teflon block, highly reflective to VUV light, is placed in front of each array of SiPMs. The block presents small holes filled with liquid xenon in front of each sensor, of the same transversal dimensions of the SiPMs and with a depth of 5mm (see figure \ref{fig:sketch}). As a consequence, when the 511 keV gammas emitted by the positron annihilation interact with the volume of xenon, the scintillation light emitted is reached in the SiPM in front of the hole.

The scintillation light detected in the SiPMs is digitized with two TOFPET2 ASICs~\cite{h, i} from the PETsys company. It is a low-power and low-noise ASIC implemented for time of flight applications. The ASIC has 64 independent channels which give the charge and time information. It is used with a dual threshold trigger with fast dark count rejection~\cite{j}. The lower threshold is used to measure the event timestamp and the higher one to reject dark counts and is used to integrate the charge to provide the energy measurement. The events that cross both thresholds are accepted and digitized, otherwise, the event is rejected without any logic dead time. Both thresholds can be configured independently in each of the 64 channels.

\section{Results}
\begin{figure}[htbp]
\centering 
\includegraphics[width=.6\textwidth]{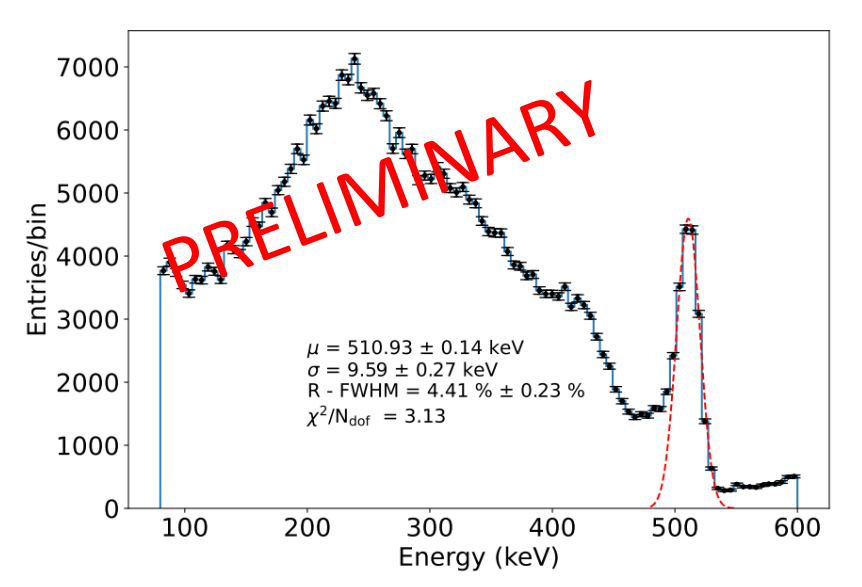}
\caption{\label{fig:ener_res} Energy spectrum obtained by one of the SiPMs, fitted to a normal function in the photo-peak.}
\end{figure}

During a run, when the charge recorded in a specific channel crosses both thresholds, the timestamp and the integrated charge of the channel are saved in a file. The channels are divided event by event according to their timestamp and those events which do not have data in both SiPM planes are rejected. Figure \ref{fig:ener_res} shows the energy spectrum obtained together with the photoelectric peak fit for one channel. 

\begin{figure}[htbp]
\centering 
\includegraphics[width=.50\textwidth]{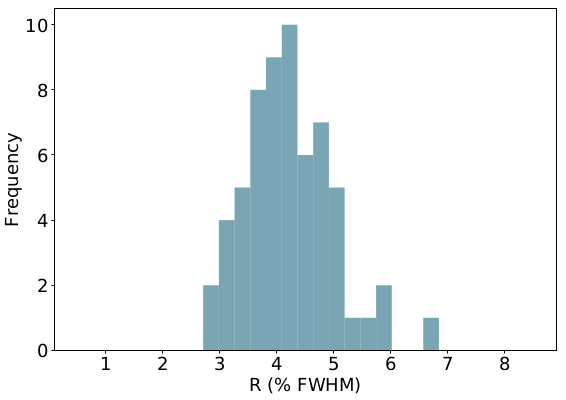}
\caption{\label{fig:all_res} Distribution of the energy resolution in each sensor.}
\end{figure}

Figure \ref{fig:all_res} shows the distribution of the energy resolution obtained for the SiPMs of one plane. The reason for choosing only one plane is because one teflon block is aligned better with the sensors than the other one, thus obtaining a more uniform response of the sensors. 

\begin{figure}[htbp]
\centering 
\includegraphics[width=.6\textwidth]{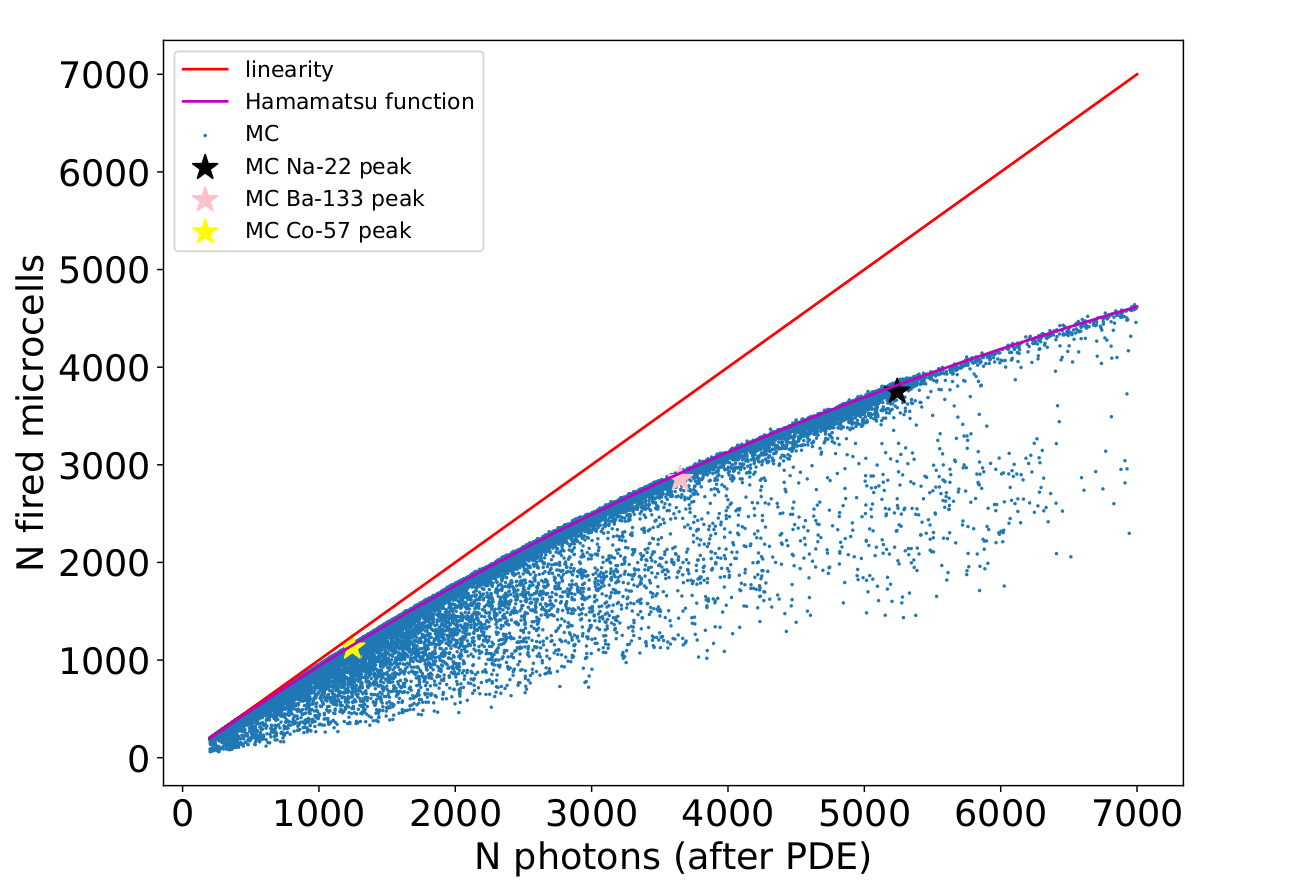}
\caption{\label{fig:sat_curve} Number of fired microcells versus the number of predicted photoelectrons (blue points). The position of the photoelectric peaks of the three calibration sources and the saturation curve provided by Hamamatsu are also shown.}
\end{figure}

The average of the energy resolution obtained is $4.2\% \pm 0.2\%$ FWHM. This result is lower than the one predicted by the Monte Carlo, which is $5.6\% \pm 0.9\%$ FWHM. This is due to the fact that the SiPMs we are using have 6162 microcells and the Monte Carlo predicts that in the photo-peak 5240 photoelectrons are detected on average. Since the number of microcells is similar to the number of detected photoelectrons, the energy measurement is likely affected by saturation, which produces a narrower peak as a result. In order to confirm this, we ran Monte Carlo simulations, where the behaviour of each SiPM microcell is described, taking into account its recovery time. We simulated three different calibration sources to investigate the SiPM response as a function of the energy. Figure \ref{fig:sat_curve} shows the number of fired microcells versus the number of photons that should be detected in the case of no saturation. Moreover, the number of photoelectrons detected in the photo-peak is shown for each calibration source together with the saturation curve provided by Hamamatsu. The calibration sources used, $^{22}$Na, $^{57}$Co and $^{133}$Ba, lie close to this saturation curve, which is far from the linearity, thus confirming the saturation in our measurement. 

\begin{figure}[htbp]
\centering 
\includegraphics[width=.50\textwidth]{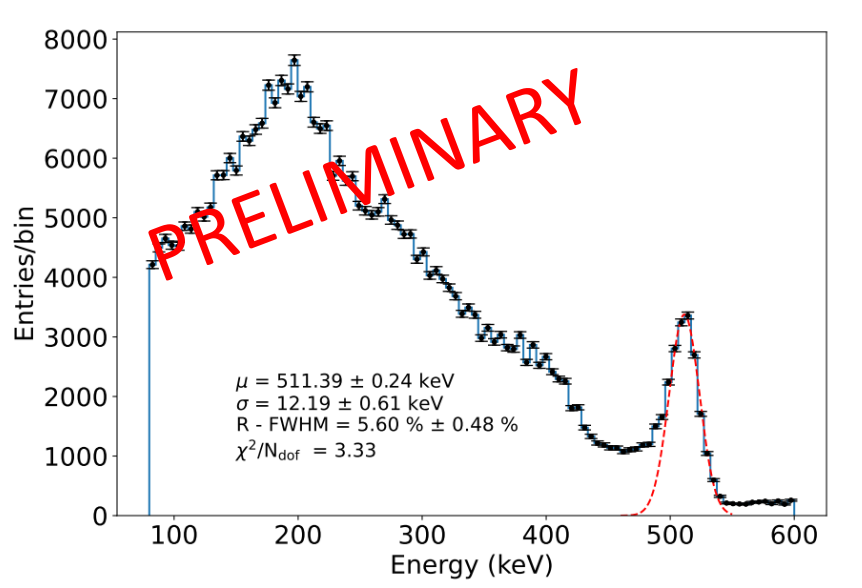
}
\qquad
\includegraphics[width=.44\textwidth]{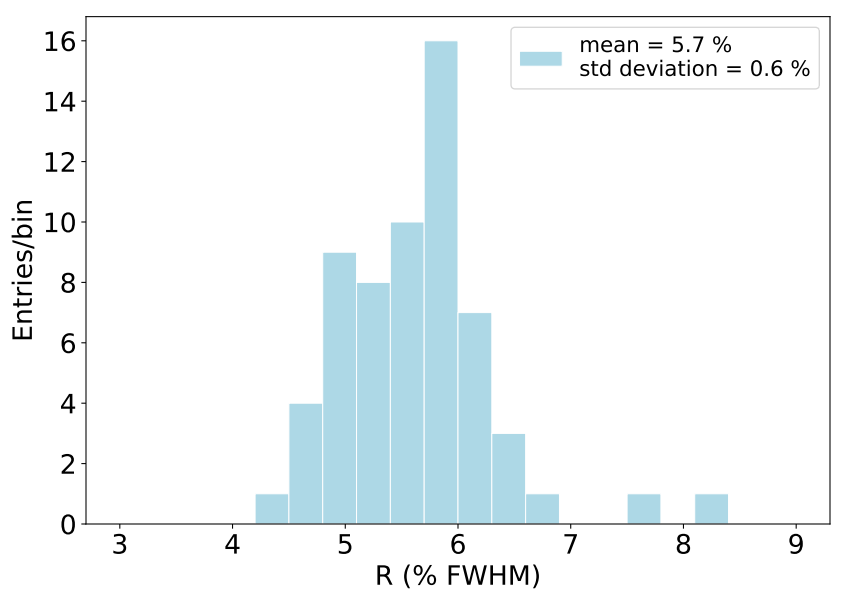}
\caption{\label{fig:ener_res_corr} Results obtained after correcting the data for saturation. Left: Energy spectrum obtained by one of the SiPMs, fitted to a normal function in the photo-peak. Right: Distribution of the energy resolution in each sensor.}
\end{figure}

Therefore, correcting the data with this saturation curve will give an energy spectrum without saturation and so a reliable result of the energy resolution. This new spectrum is shown in figure \ref{fig:ener_res_corr} together with the distribution of the energy resolution for all sensors. The energy resolution results in $5.7\% \pm 0.6\%$ FWHM, which is in agreement with the result obtained with the Monte Carlo.

Nevertheless, a new setup is being built substituting the SiPMs from the Hamamatsu company with ones from the FBK company to perform a direct measurement in data. These latter SiPMs have 5 times more microcells per area than the former, thus allowing one to obtain a measurement of the energy resolution without saturation. This will also allow us to validate our Monte Carlo simulations, comparing the results obtained.

\section{Conclusions}
The first prototype of PETALO, PETit, has been built and it is operating at Instituto de Física Corpuscular (IFIC) in Valencia, Spain. A very good energy resolution has been obtained, although the result was affected by saturation. In order to obtain a final result without saturation new SiPMs with more microcells are going to be used. In next steps, studies of the time resolution are going to be carried out. In the future, more realistic configurations of the detector will be studied, with larger xenon volumes in front of the SiPMs.

\acknowledgments

This work was supported by the European Research Council under Grant ID 757829 and is part of the PRE2021-097277 grant, funded by MCIN/AEI/10.13039/501100011033 and the ESF+.

\end{document}